\documentstyle[12pt]{article}\textheight 200mm \textwidth 160mm
\begin{document}
\setcounter{page}{0}
\title{\bf An operator approach to BRST invariant transition amplitudes}
\author{Silvio J. Rabello\thanks{e-mail: goldegol@vms1.nce.ufrj.br.bitnet}
\,\, and Arvind N. Vaidya\thanks{e-mail: ift10034@ufrj.bitnet}\\
\\{\it Instituto de F{\'\i}sica}\\
{\it Universidade Federal do Rio de Janeiro}\\
{\it Rio de Janeiro  RJ}\\
{\it Caixa Postal 68.528-CEP 21945-970}\\
{\it Brasil}}
\date{March 94}
\maketitle
\begin{abstract}

{\sl The transition amplitudes for the free spinless and  spinning
 relativistic particles are obtained by applying an operator method
 developed long ago by Dirac and Schwinger to the BFV form of the BRST
 theory for constrained systems.}

\end{abstract}

\newpage

The BRST method of quantization \cite {BRST,BV,HeTe}
is now  thought as the most effective way of dealing with constrained
dynamical systems. In trading the local invariance of gauge systems for
the symmetry under global nilpotent transformations generated by the
BRST charge, it opened the pathway to the quantum description of many
interesting systems ,e.g. the first quantized particles, strings and
membranes not to mention a better understanding of the Yang-Mills theory.

Recently the BRST invariant transition amplitudes for the spinless and
spinning relativistic \cite {Batlle} and nonrelativistic \cite {Gomis}
particles were obtained using the  BFV path integral formulation \cite{BV}.
Although the early developments of the BRST theory were done in a path
integral environment, the operator method often helps to clarify some
aspects unforeseen by the functional approach \cite{Marne}. In this paper
we obtain the transition amplitudes for the free spinless and spin one
half  relativistic particles in the BFV-BRST framework by applying
an operator approach developed long ago by Dirac \cite {Di} in his
investigations on the role of the classical action in quantum mechanics and
Schwinger \cite {Schw} in his early calculations of effective actions.

Let us start by considering a hamiltonian system with finite degrees of
freedom and linearly independent first class constraints $G_a$
$(a=1,...,k)$, that we choose to be bosonic and real for simplicity and
with the corresponding quantum operators generating the algebra
(with $U_{ab}^{\;\;c}$ being the structure constants):
\begin{equation}
\label{const}
[G_a, G_b]_{-}=iU_{ab}^{\;\;c}G_c\,\,.
\end{equation}

In the BFV-BRST theory \cite{BV} the original set of coordinate and
momentum operators $({\bf q},{\bf p})$ is extended as
\begin{equation}
\label{Space}
({\bf q},{\bf p})\rightarrow ({\bf q},{\bf p})\oplus(\lambda,\pi)
\oplus(\eta,{\cal P})\oplus(\bar\eta,\bar{\cal P})\,,
\end{equation}
where we introduced the bosonic Lagrange multipliers $\lambda^a$ and
respective momenta $\pi_a$, a fermionic  pair of conjugate ghost
operators $(\eta^a\,,\,{\cal P}_a)$ for each constraint $G_a$,
and a pair of fermionic antighost conjugate operators
$({\bar\eta}_a\,,\,{\bar{\cal P}}_a)$ to take care of each constraint
$\pi_a$ that appear due to the ``einbein'' character of $\lambda^a$
that we assume from start. These hermitian operators generate the algebra
(the non zero part of it)
\begin{equation}
\label{GhostA}
[\lambda ^a,\pi_b]_-=i\delta^a_{\;b},\;\;\;[\eta^a, {\cal P}_b]_{+}=
[\bar{\eta}^a, \bar{\cal P}_b]_{+}=\delta^a_{\;b}\,.
\end{equation}

Annihilating the physical states, the hermitian nilpotent BRST
charge operator in the BFV formulation \cite {BV} is given by
\begin{equation}
\label{charge}
Q=G_a\eta^a-\frac{i}{2}U_{bc}^{\;\;a}{\cal P}_a
\eta^b\eta^c-\frac{i}{2}U_{ab}^{\;\;b}\eta^a +\bar{\cal P}_a\pi^a\,.
\end{equation}
Due to the nilpotency of Q the following extension of
the original BRST invariant hamiltonian $H_0$ is also invariant:
\begin{equation}
\label {ext}
H=H_0+[\Psi,Q]_{+}\,.
\end{equation}
Where $\Psi$ is an arbitrary ``gauge fermion''. And also, since
$Q\vert phys\rangle=0$, the above extension of $H_0$ can be used in
the evolution operator without changing the transition amplitudes.
In fact for generally covariant systems, like the relativistic particle
and string, where the classical $H_0\approx 0$, the choice of an
appropriate $\Psi$ will prove essential to ensure integrability.

In the following we use the extended hamiltonian $H$ and the condition
$Q\vert phys\rangle=0$ to obtain the propagator for the scalar and
spinning relativistic particles by the Dirac-Schwinger method.

To illustrate the method we consider the transition amplitude
between position eigenstates for a system with one degree of freedom
and obeying the Schr\"odinger equation:
\begin{equation}
\label {Sch}
\langle q'',t\vert q',0\rangle =\langle q''\vert e^{-iHt}\vert q'\rangle \,,
\end{equation}
where $\vert q',t\rangle$ are
eigenvectors of the position operator $q(t)$ with eigenvalue $q'$
(hereafter all operator eigenvalues will be primed) and
$\langle q'',0\vert q',0\rangle=\delta (q''-q')$. We also  choose
$[q,p]_{-}=i$, with $p$ being the conjugate momentum operator,
while the following reasoning is also valid for fermionic $q$ and $p$
\cite {RaAlVa}.

To obtain the above amplitude Dirac \cite{Di} wrote it in the following way:

\begin{equation}
\label{expW}
\langle q'',t\vert q',0\rangle
=e^{iW(q'',q';t)}\,,
\end{equation}
where $W(q'',q';t)$ is a complex function of the end point
coordinates and time. It is easy to verify that this function is determined
by the following relations:

\begin{eqnarray}
\label{HJ}
-{\partial W(q'',q';t)\over{\partial t}}
&=&{\langle q'',t\vert  H( q(t), p(t))
\vert q',0\rangle /{\langle q'',t\vert q',0\rangle }},\\
\bigskip
\label{mom1}
{\partial W(q'',q';t)\over{\partial q''}}&=&{\langle q'',t\vert  p(t)
\vert q',0\rangle /{\langle q'',t\vert q',0\rangle }},\\
\bigskip
\label{mom2}
-{\partial W(q'',q';t)\over{\partial q'}}&=&{\langle q'',t\vert  p(0)
\vert q',0\rangle
/{\langle q'',t\vert q',0\rangle }},\\
\bigskip
\label{Norm}
W(q'',q';0)&=&-i\, ln \delta (q''-q').
\end{eqnarray}

To solve this problem Schwinger \cite{Schw} noticed that the above
equations relate the transition amplitude to the solution of the Heisenberg
equations for $ q(t)$ and $ p(t)$. If we solve for $ p(t)$ in terms of
$ q(t)$ and $ q(0)$ and insert this, in a time ordered fashion, on
(\ref{HJ})-(\ref{mom2}) we are left with a set of first order equations
to integrate. In the following we apply the above
Dirac-Schwinger method to the BFV-BRST models of the scalar and
spinning particles.

The classical mechanics of a spinless relativistic particle
with parametrized world line $X^\mu (s)$ $(\mu=0,...,3)$, $s\in [0,1]$
and momentum $P^\mu (s)$ , is described  by the vanishing hamiltonian
$H_0 =P^2-m^2\approx 0 $ \cite {HeTe2}. In the canonical quantization of
this system $X^\mu$ and $P^\mu$ become operators that generate the algebra
(the nonzero part)

\begin{equation}
\label {xp}
[X^\mu,P^\nu]_{-}=i\eta^{\mu\nu}\,,
\end{equation}
($\eta_{\mu\nu}=diag(1,-1,-1,-1)$) and $H_0$ becomes the gauge
generator $G=P^2-m^2$. The BFV-BRST charge for this system with a
single abelian generator is simply
\begin{equation}
\label {Q1}
Q=\eta G +{\bar{\cal P}}\pi\,.
\end{equation}
In order do construct an evolution operator in the parameter time s, we
rely on the BRST extended hamiltonian (\ref{ext}), a popular choice
for $\Psi$ that ensures the integrability of this system, is given by
\cite {HeTe}
\begin{equation}
\label{K1}
\Psi={\cal P}\lambda\,,
\end{equation}
with this choice the extended hamiltonian reads

\begin{equation}
\label{Ham}
 H=[\Psi,Q]_{+}=\lambda({ P}^2-m^2)+i{{\cal P}}{{\bar{\cal P}}}\,.
\end{equation}
{}From the above $H$ we have the Heisenberg equations (the dot means
derivative with respect to the parameter s):

\begin{eqnarray}
\label{Heis2}
\dot X^\mu   &=& 2\lambda P^\mu\,,\qquad\qquad  \dot P^\mu  =0\,,\\
\dot \lambda   &=& 0\,,\;\qquad\qquad\qquad  \dot \pi  =-(P^2-m^2)\,,
\\
\dot \eta   &=& {\bar{\cal P}}\,,\qquad\qquad\qquad
\dot {\cal P}  =0\,,\\\dot {\bar\eta}   &=& -{\cal P}\,,\;
\quad\qquad\qquad  \dot {\bar{\cal P}}  =0\,.
\end{eqnarray}
With the solutions
\begin{eqnarray}
\label{Sol1}
 X^\mu (s)&=& X^\mu(0)+2\lambda P^\mu s\,, \\
\label{Sol2}
\pi(s)&=&\pi(0)-(P^2-m^2)s\,,\\
\label{Sol3}
\eta(s)&=&\eta(0)+{\bar{\cal P}}s\,,\\
\label{Sol4}
{\bar\eta}(s)&=&{\bar\eta}(0)-{\cal P}s\,.
\end{eqnarray}
Note that in (\ref {Sol1}) there is an ambiguity, since we can
change $\lambda \rightarrow -\lambda$ and $P^\mu\rightarrow -P^\mu$
and end up with the same $X^\mu(s)$ \cite{HeTe2}. A remedy for this
situation is to restrain the eingenvalues of $\lambda$ to $(-\infty,0]$
or $[0,\infty)$ .

We now write  ${{\cal P}}$ and ${\bar{\cal P}}$ in terms of ${\eta}(s)
,\;{\eta}(0),\;{\bar{\eta}}(s)$ and ${\bar{\eta}}(0)$, so that we get
the time ordered hamiltonian :
\begin{eqnarray}
\label {Hord1}
H_{ord}&=&\lambda (P^2-m^2)-
{i\over s^2}\biggl({\bar\eta}(s)\eta(s)-
{\bar\eta}(s)\eta(0)\nonumber\\
&+&{\bar\eta}(0)\eta(0)+\eta(s){\bar\eta}(0)
-[{\bar\eta}(0),\eta(s)]_{+}\biggr)\,,
\end{eqnarray}
with the anticommutator being
\begin{equation}
\label{com}
[{{\bar\eta}}(0),{\eta}(s)]_{+}=s\,.
\end{equation}

We are now in position to integrate (\ref{HJ}),
using the basis $\vert P',\lambda',\eta',{\bar\eta}',s\rangle $
(with the ghost eigenvalues being Grassmann numbers):

\begin{equation}
\label {W1}
W=-s\lambda'(P'^2-m^2)-{i\over s}({\bar\eta}''-{\bar\eta}')
(\eta ''-\eta') -i\,ln\,s +\Phi\,,
\end{equation}
where $\Phi$ is a s independent function of the dynamical variables.
Using (\ref{mom1}), (\ref{mom2}) and the solutions (\ref {Sol1})-(\ref{Sol4})
we have
\begin{eqnarray}
\label{Phi1}
\biggl({\partial\over{\partial P_{\mu}''}}+
{\partial\over{\partial P_{\mu}'}}\biggr)\Phi=
\biggl({\partial\over{\partial \lambda''}}+
{\partial\over{\partial \lambda'}}\biggr)\Phi=0\,,\nonumber\\
{\partial\over{\partial \eta'}}\Phi={\partial\over{\partial \eta''}}\Phi=
{\partial\over{\partial {\bar\eta'}}}\Phi=
{\partial\over{\partial {\bar\eta''}}}\Phi=0\,,
\end{eqnarray}
so that with aid of (\ref {Norm}) and using the Berezin definition for
a Grassmann $\delta-$function \cite{DeW}:
\begin{equation}
\label {Phi2}
\Phi=\Phi(P_{\mu}''-P_{\mu}'\,;\lambda''-\lambda')=-i\,ln\,i\delta^4
(P''-P')\delta(\lambda''-\lambda')\,.
\end{equation}

Now that we have found $W$ we must impose the invariance under Q:
\begin{equation}
\label{Q|>=0}
Q\vert P',\pi',\eta',{\bar\eta}',0\rangle=0\,,\qquad
Q\vert P'',\pi'',\eta'',{\bar\eta}'',s\rangle=0\,,
\end{equation}
where we changed our basis from $\lambda$ to $\pi$ eigenvectors for
later convenience, and used the fact that $Q^\dagger=Q$.
Among the several ways for this to be true \cite {HeTe}, we choose
the boundary conditions:
\begin{equation}
\label {Bound}
\pi''=\pi'=\eta''=\eta'={\bar\eta''}={\bar\eta'}=0\,.
\end{equation}
The condition on ${\bar\eta}$ is a consistency one for $\pi=[Q,
{\bar\eta}]_{+}$ and $Q$ to annihilate $\vert phys\rangle$.

We now Fourier transform $e^{iW}$ to the above basis choosing $\lambda\geq0$,
\begin{equation}
\label {SchwRep1}
K(P'',P')=i \int_0^{\infty} d\lambda'\; s e^{i\lambda'(\pi''-\pi')}
e^{-i[s\lambda'(P'^2-m^2)+{i\over s}
({\bar\eta}''-{\bar\eta}')(\eta ''-\eta')]}\delta^4(P''-P')\,.
\end{equation}
With the BRST invariant boundary conditions and defining $s\lambda'\equiv T$:
\begin{equation}
\label {SchwRep2}
K(P'',P')=i\int_{0}^{\infty}dT\; e^{-iT(P'^2-m^2)}\delta^4(P''-P')=
{\delta^4(P''-P')\over{P'^2-m^2-i\epsilon}}\,.
\end{equation}
The choice of positive $\lambda$ led to the momentum space
Feynman propagator, had we chosen negative values for $\lambda$ we
would end up with the complex conjugate of it. We now turn to the more
involved model of a spin one half relativistic particle.

The pseudoclassical behavior of a  spinning relativistic particle is
described by an action where the local diffeomorfism invariance is
generalized to a local world line supersymmetry \cite {Brink}. From this
model we have five second class and four first class constraints, that
after applying the Dirac algorithm reduce to the two first class
constraints \cite{Batlle}:

\begin{equation}
\label {Constr2}
G_{1}=P^2-m^2\,,\qquad G_{2}=\zeta\cdot P-\zeta_5 m\,,
\end{equation}
where the $\zeta^\mu(s)$ and $\zeta_5(s)$ are Grassmann odd variables
describing the spin degrees of freedom (spin one half). In the quantum
version of the above model the $\zeta(s)$'s obey
\begin{equation}
\label {Commut2}
[\zeta_{\mu},\zeta_{\nu}]_{+}=-2\eta_{\mu\nu}\,,\qquad\qquad
[\zeta_5,\zeta_5]_{+}=2
\end{equation}
and the gauge algebra is
\begin{equation}
\label {Lie}
[G_{2},G_{2}]_{+}=-2(P^2-m^2)=-2G_{1}\,.
\end{equation}

As in the spinless case the original phase space will be extended with
the same variables for the gauge generator $G_1$ and the additional ones
for $G_2$, with reversed statistics, obeying
\begin{equation}
\label {Commut3}
[\lambda_{2},\pi_{2}]_{+}=1\,,\quad [\eta_{2},{\cal P}_{2}]_{-}=
[{\bar\eta_{2}},{\bar{\cal P}}_{2}]_{-}=i\,,
\end{equation}

{}From the gauge algebra we have that the single surviving
structure constant is $U_{22}^{\;\;1}=2i$, so that the BRST charge
for this model reads
\begin{equation}
\label{Charge2}
Q=\eta_{1} (P^2-m^2) + \eta_{2}(\zeta\cdot P-\zeta_5 m) + {\cal P}_{1}
{\eta_{2}}^2 +{\bar{\cal P}}_{1}\pi_{1} + {\bar{\cal P}}_{2}\pi_{2}\,.
\end{equation}
The generalization of the last section gauge fermion is
\begin{equation}
\label {Gfermion}
\Psi={\cal P}_{a}\lambda^{a}\,,
\end{equation}
leading to the extended hamiltonian
\begin{equation}
\label {Ham2}
H=[Q,\Psi]_{+}= \lambda_{1} (P^2-m^2) -i\lambda_{2}(\zeta\cdot P-\zeta_5 m)
- 2i\eta_{2}\lambda_{2}{\cal P}_{1} + i{\cal P}_{1}{\bar{\cal P}}_{1} +
{\cal P}_{2}{\bar{\cal P}}_{2}\,.
\end{equation}

The Heisenberg equations of motion are:
\begin{eqnarray}
\label{Heisf1}
\dot  X^\mu  &=&2\lambda_{1} P^\mu -i\lambda_{2}\zeta^\mu
\,,\;\qquad\quad \dot P^\mu  =0
\,,\\
\label{Heisf2}
\dot \zeta^\mu  &=&-2\lambda_{2}P^\mu\,,\qquad\qquad\qquad\;\,
\dot \zeta_5  =-2\lambda_2m
\,,\\
\label{Heisf3}
\dot \lambda_{1}  &=&0\,\,,\;\quad\qquad\qquad\qquad\qquad
\dot \pi_{1}  = -(P^2-m^2)
\,,\\
\label{Heisf4}
\dot \lambda_{2}  &=&0\,,\;\quad\qquad\qquad\qquad\qquad
\dot \pi_{2}  =-(\zeta\cdot P-\zeta_5 m) - 2\eta_{2}{\cal P}_{1}
\,,\\
\label{Heisf5}
\dot \eta_{1}  &=&{\bar{\cal P}}_{1}+2\eta_{2}\lambda_{2}\,,\qquad\qquad
\quad\dot {\cal P}_{1}  =0
\,,\\
\label{Heisf6}
\dot {\bar\eta}_{1}  &=& -{\cal P}_{1} \,,\qquad\qquad\qquad\qquad
\dot {\bar{\cal P}}_{1}  =0
\,,\\
\label{Heisf7}
\dot \eta_{2}  &=&{\bar{\cal P}}_{2} \,,\,\;\;\qquad\qquad\qquad\qquad
\dot {\cal P}_{2}  =2i\lambda_{2}{\cal P}_{1}
\,,\\
\label{Heisf8}
\dot {\bar\eta}_{2}  &=&{\cal P}_{2} \,,\,\;\;\qquad\qquad\qquad\qquad
\dot {\bar{\cal P}}_{2}  =0\,,
\end{eqnarray}
with the solutions
\begin{eqnarray}
\label {Solf1}
X^\mu (s)&=& X^\mu (0) + (2\lambda_{1}P^\mu -i\lambda_{2}\zeta^\mu(0))s \,,\\
\label {Solf2}
\zeta^\mu (s)&=&\zeta^\mu (0)-2\lambda_{2}P^\mu s \,,\\
\label {Solf2.1}
\zeta_5 (s)&=&\zeta_5(0)-2\lambda_2 ms\,, \\
\label {Solf3}
\pi_{1}(s)&=&\pi_{1}(0)-(P^2-m^2)s\,,\\
\label {Solf4}
\pi_{2} (s)&=&\pi_{2}(0) - (2\eta_{2}(0){\cal P}_{1} +\zeta (0)\cdot P
-\zeta_5(0)m)s+[\lambda_{2}(P^2-m^2)-{\bar{\cal P}}_{2}{\cal P}_{1}]s^2
\,,\\
\label {Solf5}
\eta_{1} (s)&=& \eta_{1} (0) + ({\bar{\cal P}}_{1} -
2\lambda_{2}\eta_{2}(0))s -\lambda_{2}{\bar{\cal P}}_{2}s^2\,,\\
\label {Solf6}
{\bar\eta}_{1}(s)&=&{\bar\eta}_{1}(0)-{\cal P}_{1}s\,,\\
\label {Solf7}
\eta_{2}(s)&=& \eta_{2}(0)+{\bar{\cal P}}_{2}s\,,\\
\label {Solf8}
{\bar\eta}_{2}(s)&=&{\bar\eta}_{2}(0)+{\cal P}_{2}(0)s
-i\lambda_{2}{\cal P}_{1}s^2\,.
\end{eqnarray}
We can see that due to the nilpotency of $\lambda_2$ we can replace
$\zeta^\mu(s)$ and $\zeta_5(s)$ by  $\zeta^\mu(0)$ and $\zeta_5(0)$
in $H$ without changing anything. Choosing the operator initial
conditions $\zeta^\mu(0)=\gamma_5\gamma^\mu$ and $\zeta_5(0)=\gamma_5$,
with $\gamma^\mu$ and $\gamma_5$ being the Dirac matrices, we get closer to
the usual description of field theoretical fermions. As in the spinless
case we use the above solutions to obtain the time ordered hamiltonian:
\begin{eqnarray}
\label {Hord2}
H_{ord}&=&\lambda_{1} (P^2-m^2) -i\lambda_{2}\gamma_5(\gamma\cdot P-m)-
{1\over s^2}\biggl[i\biggl({\bar\eta_{1}}(s)\eta_{1}(s)-
{\bar\eta_{1}}(s)\eta_{1}(0)\nonumber\\
&+&{\bar\eta_{1}}(0)\eta_{1}(0)+\eta_{1}(s){\bar\eta_{1}}(0)
-[{{\bar\eta}}_{1}(0),{\eta}_{1}(s)]_{+}\biggr)
-{\bar\eta_{2}}(s)\eta_{2}(s)\nonumber\\
&+&{\bar\eta_{2}}(s)\eta_{2}(0)
-{\bar\eta_{2}}(0)\eta_{1}(0)+\eta_{2}(s){\bar\eta_{2}}(0)
+[{{\bar\eta}}_{2}(0),{\eta}_{2}(s)]_{-}\biggr]\,,
\end{eqnarray}
where the (anti)commutators are
\begin{equation}
\label{com2}
[{{\bar\eta}}_{1}(0),{\eta}_{1}(s)]_{+}=s\, , \qquad
[{{\bar\eta}}_{2}(0),{\eta}_{2}(s)]_{-}=is\,.
\end{equation}
Unlike the case of a single bosonic constraint the contribution of
the fermionic ghost anticommutator exactly cancels the commutator
of the bosonic ghost in the hamiltonian. This was expected due to the
supersymmetry of the model.

Integrating for $W$ in the basis
$\vert P',\lambda',\eta',{\bar\eta}',s\rangle$ we have:
\begin{eqnarray}
\label{W3}
W=&-s\lambda_{1}'(P'^2-m^2) +is\lambda_{2}'\gamma_5(\gamma\cdot P'-m)
+{1\over s} [-i({\bar\eta_{1}}''-{\bar\eta_{1}}')(\eta_{1} ''-\eta_{1}')
\nonumber\\ &+({\bar\eta_{2}}''-{\bar\eta_{2}}')(\eta_{2} ''-\eta_{2}')]
+\Phi\,,
\end{eqnarray}
repeating the same analysis as in the previous case we have for the
s independent $\Phi$:

\begin{equation}
\label{phi}
\Phi=-i\,ln\,i\delta(\lambda_1''-\lambda_1')\delta(\lambda_2''-
\lambda_2')\delta^4(P''-P')\,.
\end{equation}

It is easy to see that the same invariant boundary conditions of the
spinless case hold:
\begin{equation}
\label{Bound2}
\pi_a''=\pi_a'=\eta_a''=\eta_a'={\bar\eta_a''}={\bar\eta_a'}=0\,.
\end{equation}

To impose the above condition on $\pi_a$ we fourier tranform $e^{iW}$
in both $\lambda_1$ $(\lambda_1\geq 0)$ and $\lambda_2$, in the Berezin
sense for $\lambda_2$ \cite {DeW}. Defining $s\lambda_1'\equiv T$
and $s\lambda_2'\equiv \theta$ we get
\begin{eqnarray}
\label {SchwRep4}
K(P'',P')&=&i\int_{0}^{\infty}dT\; e^{-iT (P'^2-m^2)}\int d\theta
e^{-\theta\gamma_5(\gamma\cdot P'-m)}\delta^4(P''-P')\nonumber\\
&=&-\gamma_5{\gamma\cdot P'-m\over{P'^2-m^2-i\epsilon}}\delta^4(P''-P')\,.
\end{eqnarray}
That is the momentum space Feynman propagator for the
Dirac equation, modulo a $-\gamma_5$ factor that is unavoidable in these
supersymmetric models of spinning particles \cite{Batlle,HeTe2}.

\bigskip
\noindent{\Large\bf Discussion }
\bigskip

In this article we obtained the BRST invariant transition amplitudes for
relativistic particles in the operator framework provided by the
Dirac-Schwinger method. It remains a challenge to apply this procedure
to more involved first quantized problems like the higher spin point
particles and the spinning string.

The authors are grateful to Patricio A. Gaete Dur\'an and Carlos Farina de
Souza for reading the manuscript and for many stimulating discussions.
This work was partially supported by CNPq.

\end{document}